\documentclass[12pt]{article}
\usepackage{amsmath,amssymb,amscd,cite}
\newtheorem{thm}{Theorem}[section]
\newtheorem{prop}[thm]{Proposition}
\newtheorem{lem}[thm]{Lemma}
\newtheorem{cor}[thm]{Corollary}
\newtheorem{defi}[thm]{Definition}
\newcommand{\pf}{{\bf Proof. \ }}
\newcommand{\qed}{\hfill $\blacksquare$ \\}
\font\msbm=msbm10 at 12pt
\newcommand{\Z}{\mbox{\msbm Z}}
\newcommand{\ZZ}{\mbox{\msbm Z}}

\newcommand{\FF}{\mbox{\msbm F}}
\newcommand{\F}{\mbox{\msbm F}}
\newtheorem{rem}[thm]{Remark}
\newtheorem{ex}[thm]{Example}
\newcommand{\vv}{\mathbf{v}}

\newcommand{\ord}{ord}
\begin{document}
\title{Constacyclic  Codes Over Finite\\Principal Ideal Rings}
\author{Aicha Batoul, Kenza Guenda and T. Aaron Gulliver
\thanks{A. Batoul and K. Guenda
are with the Faculty of Mathematics USTHB, University
of Science and Technology of Algiers, Algeria. T. Aaron Gulliver is
with the Department of Electrical and Computer Engineering,
University of Victoria, PO Box 3055, STN CSC, Victoria, BC, Canada
V8W 3P6. email: agullive@ece.uvic.ca.}}
\date{}
\maketitle
\begin{abstract}
In this paper, we give an important isomorphism between contacyclic codes and cyclic codes,over finite
principal ideal rings.Necessary and sufficient conditions for the existence of non-trivial cyclic self-dual
codes over finite principal ideal rings are given.
\end{abstract}

\section{Introduction}

Although codes over rings are not new~\cite{blake}, they have
attracted significant attention from the scientific community only
since 1994, when Hammons et al.~\cite{sole1} established a
fundamental connection between non-linear binary codes and linear
codes over $\ZZ_4$. In~\cite{sole1}, it was proven that some of the
best non-linear codes, such as the Kerdock, Preparata, and Goethal
codes can be viewed as linear codes over $\ZZ_4$ via the Gray map
from $\ZZ_4^n$ to $\FF_2^{2n}$. The link between self-dual codes and
unimodular lattices was given by Bonnecaze et al.~\cite{bonnecaze}
and Bannai et al.~\cite{bannai}. These results created a great deal
of interest in self-dual codes over a variety of rings, see
\cite{RS} and the references therein. Dougherty et
al.~\cite{CRT,dkk} used the Chinese remainder theorem to generalize
the structure of codes over principal ideal rings. They gave
conditions on the existence of self-dual codes over principal ideal
rings in~\cite{CRT}.

Dougherty~\cite{steven} recently posed a number of problems
concerning codes over rings. Several of these are answered in this
paper. In particular, we give necessary and sufficient conditions on
the existence of self-dual codes over principal ideal rings. The existence
of such codes requires the existence of self-dual codes over all the base
finite chain ring. We also give the structure of constacyclic codes over
finite principal ideal rings. The projection and the lift of these
codes is described using a generalization of the Hensel Lift Lemma
and the structure of the ideals of $R[x]/\langle x^n-\lambda
\rangle$. Finally, infinite families of self-dual codes are
given over principal ideal rings.
Codes over rings are a generalization of codes over fields.
In~\cite{sole1}, it was proven that some well known non-linear
codes, such as the Kerdock, Preparata, and Goethal codes, are the
image of linear codes over $\ZZ_4$ via the Gray map from $\ZZ_4^n$
to $\FF_2^{2n}$. These results generated a great deal of interest in
self-dual codes over a variety of rings, e.g. \cite{RS,G-G,D-L}. In
addition to self-dual codes over rings being theoretically
important, they also have many practical applications, for example
they are related to unimodular lattices~\cite{bannai,bonnecaze}.
Several researchers have considered self-dual codes over
rings~\cite{dgw,doughertyself,D-L}. The structure of cyclic codes
over $\ZZ_{p^a}$ was first given by Calderbank and Sloane~\cite{CS}.
This motivated others (e.g.~\cite{abualrub,blackford,wolfmann}), to
investigate the structure of cyclic and negacyclic codes over chain
rings. Kanwar, Dinh and L\'{o}pez-Permouth~\cite{K-L,permounth}
generalized this structure to cyclic and negacyclic codes over
finite chain rings, and considered the self-duality of these codes.
More recently, the structure given in~\cite{permounth} has been
generalized to constacyclic codes~\cite{G-G}. Motivated by an open
question posed by Jia et al.~\cite{jia} on the structure of cyclic
self-dual codes over rings, we give in this paper necessary and
sufficient conditions on the existence of non-trivial cyclic
self-dual codes over finite chain rings. Another motivation of the
present work is the characterization of those integers $n$ for which
$p^i\neq -1 \bmod n$ for all $i$ and $p$ odd. This is required to
determine the non-trivial cyclic self-dual codes given by Dinh and
L\'{o}pez-Permouth~\cite[p. 1734]{permounth}. We prove that in the
case of even nilpotency, there exists a non-trivial cyclic self-dual
code of length $n$ over a finite principal ideal ring $R$, such that the
residual field has cardinality $p^r$, if and only if $\ord_n(p^r)$
is odd. We also prove that there are no free cyclic self-dual codes
over finite chain rings with odd characteristic. Furthermore, it is
proven that a self-dual code over a chain ring cannot be the lift of
a binary cyclic self-dual code. We give explicit expressions for the
number of cyclic self-dual codes over chain rings and provide
examples.

\section{Preliminaries}
Since principal ideal rings are Frobenius rings we need to give some tools necessaries
for the after.
\section{Commutative Frobenius rings}
We assume that all rings are commutative and with identity.
For all unexplained terminology and more detailed we refer to \cite{Matsumura} (related algebra)
and to \cite{Mac}
A finite commutative ring $R$  and is Frobenius if the $R$-module $R$ is injective.
Alternatively,we can say a finite commutative ring is Frobenius if $R/J(R)$ is isomorphic
to Soc($R$),where  $R/J(R)$ is the Jacobson radical of the ring $R$ and Soc($R$) is the
Socle of the ring. Recall that the Jacobson radical is the intersection of all maximal ideals
in the ring and the Socle of the ring is the sum of the minimal $R$-submodules.
Finite Frobenius rings are very important in coding theory for several reason
and precisely for the following equality \cite{wood1}:
A code $C$ over a finite Frobenius ring $R$ and its dual satisfy the following
\begin{equation}
 \label{wood} |C||C^{\bot}|=|R|^n, \text{ and
}(C^{\bot})^{\bot}=C.
\end{equation}

If $I$ is an ideal of a finite ring,then the chain $I\supset I^2\supset I^3\supset\cdots $ stabilizes.
The smallest $e\geq1$ such that $I^e=I^{e+1}=\cdots$ is called the index of stability of $I$.
If $I$ is nilpotent,then the smallest $e\geq1$ such that $I^e=0$ is called the index of nilpotency of
$I$ and is the same as the index of stability of $I$.
Note that if $R$ is local,with maximal ideal $M$ then we have necessarily $M^e=M^{e+1}=\cdots={0}$.
Thus in the case of finite local rings, the index of stability of $M$ is in fact the index of  nilpotency of $M$.
On the other side,if $R$ has at least two maximal ideals, then for any maximal ideal $J$,$J^e=J^{e+1}=\cdots \neq{0}$.
Otherwise,if $I\neq J$ is another maximal ideal,we would have $I\supset(0)=J^e$, hence $J\subset I$, a contradiction.

Let $R$ be a ring,$I$ an ideal of $R$.
Denote by $\Psi_i:R\longrightarrow R/I$ the canonical homomorphism $x\longmapsto x+I$.
If $n$ is a fixed positive integer we also denote  $\Psi_i:R^n\longrightarrow (R/I)^n$
the canonical $R$-linear map
\[(x_1,\ldots,x_n)\longmapsto (x_1+I,\ldots,x_n+I)\]

Let $R$ be a finite ring,

Let $\mathfrak{m}_1,\mathfrak{m}_2,\dots,\mathfrak{m}_k$ the maximal  ideals of $R$
$e_1,\ldots,e_k$ their indices of stability.
Then the ideals $\mathfrak{m}_1^{e_1},\mathfrak{m}_2,\dots,\mathfrak{m}_k^{e_k}$ are  relatively prime in pairs,
 and  $\prod_{i=1}^k\mathfrak{m}_i^{e_i}=\cap_{i=1}^n \mathfrak{m}_i^{e_i}=\{0\}$.
 By the ring version of the Chinese Remainder Theorem, the canonical ring homomorphism
 \[ \Psi:R \longrightarrow \prod_{i=1}^k R/\mathfrak{m}_i^{e_i}\]
 defined by $x\longmapsto (x+\mathfrak{m}_1^{e_1},\ldots,x+\mathfrak{m}_k^{e_k})$,
 is an isomorphism.
 Denote the local rings $R/\mathfrak{m}_i^{e_i}$ by $R_i$ $(i=1,\ldots,k)$.
 The maximal ideal of $R_i$ has nilpotency index $e_i$.
 Note that $R$ is Frobenius if and only if each $R_i$ is Frobenius \cite{wood1}.
 For a code $C\subset R^n$ over $R$ and the maximal ideal $ \mathfrak{m}_i$ of $R$,
 the $\mathfrak{m}_i$-projection of $C$ is defined by $C_i=\Psi_i(C)$
 where  $\Psi_i:R^n\longrightarrow R_i^n$ is the canonical map.
 We denote by $\Psi:R^n\longrightarrow \prod_{i=1}^k R_i^n$ the map
 defined by $\Psi(u)=(\Psi_1(u),\ldots,\Psi_k(u))$ for $u \in R^n$.
 By the module version of the Chinese Remainder Theorem, the map $\Psi$ is an $R$-module isomorphism
 and
 \[ C\simeq C_1\times C_2\times\cdots C_k\]
Conversely,given codes $C_i$ of length $n$ over $R_i$  $(i=1,\ldots,k)$, we define the code
$C=CRT(C_1,\ldots,C_k)$ of length $n$ over $R$ in the following way

\[
\begin{array}{ccl}
C &=&\{ \Psi^{-1}(u_1,\ldots,u_k);\,\,u_i\in C_i\,(i=1,\ldots,k)\} \\
&=& \{u\in R^n; \,\,\Psi_i(u)\in C_i\,(i=1,\ldots,k)\}\\
\end{array}
\]
then the code $C=CRT(C_1,\ldots,C_k)$ is called the Chinese product of the code $C_i$.

As a particular case of the above discussion (and with the above notation)
we have
\begin{thm}
Let $R$ be a finite Frobenius ring,$n$ a positive integer.Then
\[ R^n=CRT(R_1^n,R_2^n,\ldots,R_k^n)\]
where $R_i$ is local Frobenius ring.
\end{thm}

\begin{lem}
Let $ C_1,C_2,\ldots,C_k$ be codes of length $n$ with $C_i$ a code over $R_i$, and let $C=CRT(C_1,C_2,\ldots,C_k)$
then
\begin{itemize}
\item[(i)]  $|C|= \prod_{i=1}^k |C_i|$.
\item[(ii)] $C$ is a free code if and only if each $C_i$ is a free code of the same rank.
\end{itemize}
\end{lem}
Notice that if two codes are free but not of the same rank then the cardinality of their
image under CRT is not that of free code.
For example,the Chinese product of free code of rank 1 over $\mathbb{Z}_2$ and a code of rank
2 over $\mathbb{Z}_3$ has cardinality $2^1\times 3^2=18$ which is not $6^k$ for any integer $k$.

\subsection{Finite Principal Ideal Rings}

\begin{lem}(\cite{Bourbaki_CA}, p.~54, Proposition 6)
\label{crtm}
Let $\mathfrak{a}_1,\mathfrak{a}_2,\dots,\mathfrak{a}_n$ be ideals
of $R$, relatively prime in pairs, and let $\mathfrak{a}=\cap_{i=1}^n
\mathfrak{a}_i$. For every $R$-module $M$, the canonical
homomorphism $M\to \prod_{i=1}^n (M/\mathfrak{a}_iM)$ is  surjective
and has kernel $\mathfrak{a}M$.
\end{lem}
Let $\mathfrak{a}_i$ be an ideal of a ring $R$, and denote $R_i =
R/\mathfrak{a}_i$. Hence we have a canonical epimorphism $\psi _i :
R \to R_i$.

\begin{prop} \label{pirs}
Let $R$ be a finite commutative ring. Then the following are
equivalent.
\begin{itemize}
\item[(i)] $R$ is a principal ideal ring.
\item[(ii)] $R$ is isomorphic to a finite product of chain rings.
\end{itemize}
Moreover, the decomposition in (ii) is unique up to the
order of factors. It has the form $R\cong \prod_{i=1}^k R/\mathfrak
m_i^{t_i}$, where $\mathfrak m_1, \mathfrak m_2,\dots, \mathfrak
m_k$ are maximal ideals of $R$, and $t_1, t_2, \dots, t_k$ are the
corresponding indexes of stability.
\end{prop}
\medskip
If $R$ is a finite principal ideal ring, we say that the
decomposition of $R$ into a product of finite chain rings, as in
(ii), is a \textit{canonical decomposition of $R$}. The ideal
$\mathfrak m_1, \mathfrak m_2,\dots , \mathfrak m_k$ in this case is
called a \textit{direct decomposition of $R$}.

\begin{lem}(\cite{bourbaki},p 110,Proposition 10)
 Let $R$ be a finite ring and $(\mathfrak{a}_i)_{i=1}^n$  be ideals of $R$. The following are equivalent:
 \begin{enumerate}
   \item [i)] The family $(\mathfrak{a}_i)_{i=1}^n$ is a direct decomposition of $R$.
   \item [ii)] For $i\neq j$, $\mathfrak{a}_i$ and $\mathfrak{a}_j$ are relatively prime and $\bigcap_{i=1}^n\mathfrak{a}_i=\{0\}$.
   \item [iii)]For $i\neq j$, $\mathfrak{a}_i$ and $\mathfrak{a}_j$ are relatively prime and $\prod_{i=1}^n\mathfrak{a}_i=\{0\}$.
   \item [iv)] There exists a family $(e_i)_{i=1}^n$ of idempotents of $R$ such that $e_ie_j\,=\,0$ for $i\neq j$,
   $1= \sum e_i$ and $\mathfrak{a_i}= R(1-e_i)$ for $i=1,\ldots,n$.
 \end{enumerate}
\end{lem}
\begin{prop}
\label{prop:doug}(\cite{dkk},Proposition 2.4)  With the notation as above,   $(\mathfrak{a}_i)_{i=1}^n$ a direct decomposition of $R$
and $M$ an $R$-module.:
\begin{enumerate}
  \item [i)] For each $i\in\{1,\ldots,n\}$ the submodule $M_i=e_iM$ is a complement of the submodule $\mathfrak{a}_iM=(1-e_i)M$
  and so the map
  \[ \psi_i :M_i\longrightarrow M/\mathfrak{a}_iM \,,\,x\longmapsto x+\mathfrak{a}_iM\]
  \item [ii)]The action of $R$ on $M$, $(r,x)\longmapsto rx$ can be identified with the componentwise actions
  $((r_1+\mathfrak{a}_1,\ldots,r_k+\mathfrak{a}_k),x_1\oplus x_2\oplus+\cdots \oplus x_n)\longmapsto r_1x_1\oplus\cdots\oplus r_nx_n$
  $((r_1+\mathfrak{a}_1,\ldots,r_k+\mathfrak{a}_k),(x_1+\mathfrak{a}_1M,\ldots x_n\mathfrak{a}_nM)\longmapsto r_1x_1+\mathfrak{a}_1M,\ldots,r_nx_n+\mathfrak{a}_nM)$
  of $\prod_{i=1}^nR/\mathfrak{a}_i $ on $M=\bigoplus_{i=1}^nM_i$ and $\prod_{i=1}^nM/\mathfrak{a}_iM $respectively
  \item [iii)]Every sub-module $N$ of $M$ is an internal direct sum of submodule $N_i=e_iN\subset M_i$ which are isomorphic
  via $\psi_i$ with the submodule $N_i'=(\mathfrak{a}_iM+e_iN)/\mathfrak{a}_iM$ of $M/\mathfrak{a}_iM$ for $i=1,2,\ldots,n$.
  Each $N_i'$ is isomorphic to $N/\mathfrak{a}_iN$ and so the decomposition $N\longrightarrow\bigoplus_{i=1}^nN_i' \subset \bigoplus_{i=1}^n M/\mathfrak{a}_iM$ canonically corresponds to the decomposition   $N\longrightarrow\bigoplus_{i=1}^n N/\mathfrak{a}_iN$
  Conversely, if for every $i=1,\ldots,n$ $N_i'$ is a submodule of $M/\mathfrak{a}_iM$ then there is a unique submodule $N=\bigoplus_{i=1}^nN_i$
  of $M$ such that $N$ is isomorphic with $\bigoplus_{i=1}^nN_i'$ via $\psi=\bigoplus_{i=1}^n\psi_i$
  \end{enumerate}
\end{prop}

Let $R$ be a finite ring. A code is a subset of $R^n$ and linear code over $R$ is an $R$-submodule of $R^n$.
In this case we say the code has length $n$.
We attach the standard inner product to the ambient space,
i.e., $[{u},{v}] = \sum u_iv_i$.
The dual code $C^\perp$ of $C$ is defined by
\begin{equation}
C^\perp=\{ {u} \in R^n \  | \ [{u},{v}]= 0 {\rm \  for\ all\ } {v}
\in C\}.
\end{equation}
We say that a code is self-orthogonal if $C \subseteq C^\perp$,
and self-dual if $C=C^\perp$.
The Hamming weight of a vector from $R^n$ is the number of non-zero coordinates in
that vector and the minimum weight is the smallest of all non-zero weights in a code.
A code $C\subset R^n$ is called a free code if $C$ is a free $R$-module,that $C$ is isomorphic to
the $R$-module $R^k$ for some $k$.

%\bigskip
%Let $R$ be a finite ring and $(\mathfrak{a}_i)_{i=1}^n$ its direct
%decomposition and let:
%\begin{equation}
%\begin{array}{ccl}
%\label{eq:doug}\psi:R^n&\longrightarrow & \prod_{i=1}^k R_i^n\\
% x &\mapsto & \psi(x) = (\psi_1(x),\psi_2(x),\ldots,\psi_k(x))
%\end{array}
%\end{equation}
%
%be the canonical $R$-module isomorphism.
%and let
%for $i=1,\dots, k$, let $C_i$ be a code over $R_i$ of length $n$ and
%let
%\[
%C=CRT(C_1,C_2,\dots,C_k)=\Psi^{-1}(C_1\times \dots \times C_k)= \{\Psi^{-1} (\vv_1,\vv_2,\dots,\vv_k) \ |\ \vv_i \in
%C_i \}.
%\]
We refer to $C$ as the \textit{Chinese product of codes $C_1,C_2,\dots,C_k$} \cite{MDR}.
\subsection{ Finite Chain Rings}
In this subsection, we summarize the necessary results
from (\cite{G-G}\cite{permounth} \cite{Ana}). A finite {\it chain ring} is a finite
commutative ring $R$ with $1 \neq 0$ such that its ideals are
ordered by inclusion. The ring $R$ is called a local ring if $R$ has
a unique maximal ideal. A finite commutative ring is a finite chain
ring if and only if it is a local principal ideal
ring~\cite[Proposition 2.1]{permounth}. Let $\mathfrak m$ be the
maximal ideal of the finite chain ring $R$. Since $R$ is principal,
there exists a generator $\gamma \in R$ of $\mathfrak m$. Then
$\gamma$ is nilpotent with nilpotency index some integer $e$. The
ideals of $R$ form the following chain
\[
<0>= \langle\gamma^e\rangle \subsetneq \langle\gamma^{e-1}\rangle
\subsetneq \ldots \subsetneq \langle\gamma\rangle \subsetneq R.
\]
The nilradical of $R$ is then $\langle\gamma\rangle$, so all
the elements of $\langle\gamma\rangle$ are nilpotent. Hence the
elements of $R\setminus \langle\gamma\rangle$ are units. Since
$\langle\gamma\rangle$ is a maximal ideal, the residue ring
$\frac{R}{\langle\gamma\rangle}$ is a field which we denote by $K$.
The natural surjective ring morphism is given by $(-)$ as follows
\begin{equation}
\label{eq:over}
\begin{split}
-:  R&  \longrightarrow K\\
a &\longmapsto \overline{a}=a \pmod \gamma
\end{split}
\end{equation}
Let $|R|$ denote the cardinality of $R$, and $R^*$ the
multiplicative group of all units in $R$. We also have that if
$|K|=q=p^r$ for some integer $r$, then
\begin{equation}
\label{cardinality-of-R} |R|=|K|\cdot|\langle\gamma\rangle|=|K|\cdot
|K|^{e-1}=|K|^e=p^{er}.
\end{equation}
We define the characteristic of the finite chain ring as the prime
number $p$ which is the characteristic of the residue field $K$ of
$R$. Note that this is not the usual definition of the
characteristic of a ring.

%A code $C$ of length $n$ over $R$ is a
%subset of $R^n$. If the code is a submodule we say that the code is
%linear. Here, all codes are assumed to be linear.
%We attach the standard inner product to the ambient space, i.e.,
%$[{v},{w}] = \sum v_iw_i$. The dual code $C^\perp$ of $C$ is defined as
%\begin{equation}
%C^\perp=\{ {v} \in R^n \  | \ [{v},{w}]= 0 {\rm \  for\ all\ } {w}
%\in C\}.
%\end{equation}
%If $C \subseteq C^\perp$, we say that the code is self-orthogonal
%and if $C=C^\perp$, we say that the code is self-dual.
A code $C$
and its dual satisfy the following
\begin{equation}
 \label{wood} |C||C^{\bot}|=q^{en}=|R|^n, \text{ and
}(C^{\bot})^{\bot}=C.
\end{equation}
\begin{rem}
\label{rem:wood} From (\ref{wood}), there exists a self-dual code of
length $n$ over $R$ if and only if $en$ is even. This explains for
example why there are no self-dual codes of odd length over
$\Z_8$~\cite{dgw}. If $e$ is even, there exists a trivial self-dual
code of length $n$ given by the generator matrix
$G=\gamma^{\frac{e}{2}}I_n.$
\end{rem}

Let $n$ be a positive integer and $q$ a prime power.
We denote by $\ord_n(q)$ the multiplicative order of $q$ modulo $n$, which is
the smallest integer $l$ such that $q^l\equiv 1 \pmod n$.

\section{Constacyclic Codes over Finite Principal Ideal Rings}
 This section considers codes over finite commutative rings which are finite
 principal ideal.
 Let $R$ be a commutative ring with unity.
For a given unit $\lambda \in R$, a code $C$ is said to be
constacyclic, or more generally, $\lambda$ constacyclic, if
$(\lambda c_{n-1}, c_0, c_1, \ldots, c_{n-2})\in C$ whenever $(c_0,
c_1,\ldots, c_{n-1}) \in C$. For example, cyclic and negacyclic
codes correspond to $\lambda=1$ and $-1$, respectively.

The main goal of this section is to prove an the existence of an isomorphism between
constacyclic codes and cyclic codes over finite principal ideal rings. This
justifies our restriction to cyclic codes in the following sections.\\
But before we recall some result given in \cite{aicha2011}

\subsection{Constacyclic Codes over Finite chain Rings}

Let $R$ be a finite chain ring,  with residue field
$\F_q$.
\begin{defi}
A polynomial $f$ of $R[x]$ is called basic irreducible if
$\overline{f}$ is irreducible in $\overline{R}[x]=[x]$. Two
polynomials $f$ and $g$ in $R[x]$ are called coprime if
$$R[x]=\langle f \rangle +\langle g \rangle.$$
Let $\lambda$ be a unit in a finite chain ring $R$. If a polynomial
$f(x)$ divides $x^{n}-\lambda$, (say $x^{n}-\lambda=f(x)g(x)$), we
refer to $g(x)=\frac{x^{n}-\lambda}{f(x)}$ as $\hat{f}(x)$.
\end{defi}
\begin{thm}(\cite[Theorem 4.7]{G-G})
\label{lem:aicha} Let $\lambda$ be a unit in a finite chain ring $R$
with characteristic $p$. When $(n,p)=1$, the polynomial
$x^n-\lambda$ factors uniquely as a product of monic basic
irreducible pairwise coprime polynomials over $R$. Furthermore,
there is a one-to-one correspondence between the set of basic
irreducible polynomial divisors of $x^n-\lambda$ in $R[x]$ and the
set of irreducible divisors of $\overline{x^n-\lambda}$ in $K$.
\end{thm}
\begin{thm}(\cite[Theorem 4.16, Corollary 4.17]{G-G})
\label{th:prince} Let $R$ be a finite chain ring and $C$ a $\lambda$
constacyclic code over $R[x]$ of length $n$ such that $(n,p)=1$,
where $p$ is the characteristic of $\overline{R}$. Then there exists
a unique family of pairwise coprime polynomials $F_0, \ldots, F_i$
in $R[x]$ such that $F_0 \ldots F_e=x^n-\lambda $ and
$C=\langle\hat{ F}_1,\gamma \hat{F}_2, \ldots,\gamma^{e-1} \hat{F}_e
\rangle$, where $\hat{F}_j=\frac{x^n-1}{F_j}$ for $0<j\le e$.
Moreover, we have that
\begin{equation}
\label{car:cons} |C|=(K)^{\sum_{j=0}^{e-1}(e-j)\deg F_{e+1}},
\end{equation}
and the ring $R[x]/\langle x^n-\lambda \rangle$ is a principal ideal
ring.
\end{thm}
It was proven by Dinh and L\'{o}pez-Permouth~\cite{permounth} that
negacyclic codes of odd length are isomorphic to cyclic codes of the
same length if $(n,p)=1$. In the following, we give an
isomorphism in more general case. For this, let $\lambda,\delta$
be a units of $R$ such that $\lambda=\delta^n$.

\begin{prop} \cite{aicha2012}
\label{prop:isom} Let $n$ be an integer and $\lambda,\delta$  units
such that $\lambda=\delta^n$. Let $\mu$ be the map $\mu :R[x]/\langle
x^n-1\rangle \mapsto R[x]/ \langle x^n- \lambda \rangle$ defined by
$\mu(c(x))=c(\delta^{-1} x)$. Then we have that $\mu$ is a ring
isomorphism.
\end{prop}
From Theorem~\ref{th:prince}, we have that the ideals in $R[x]/
\langle x^n- \lambda \rangle$ are principal ideals. Then the
following result is a straightforward consequence of
Proposition~\ref{prop:isom}.
\begin{cor}
\label{cor:square}\cite{aicha2012}
Let $R$ be a finite chain ring and $\lambda,\delta$  units in $R$ such that $\lambda=\delta^n$ . A subset
$I$ in $R[x]$ is an ideal in $R[x]/\langle x^n-1\rangle$ if and only
if $\mu(I)$ is an ideal in $R[x]/\langle x^n-\lambda \rangle$.
Equivalently, the set $C$ is a cyclic code of length $n$ over the
chain ring $R$ if and only if $\mu(C)$ is a $\lambda$-constacyclic
code of length $n$ over $R$.
\end{cor}
\subsection{Constacyclic Codes over Finite Principal Ideal Rings}
In the following we generalize the above result to finite principal ideal rings
 but before we need to recall some results about them.

For the after, we need the followings lemmas.
\begin{lem}\cite{Mac}
\label{lem:1}
Let $R^*$ denote the group of units of a finite ring $R$, if $R$ decomposes as a direct
sum $R=R_1\oplus \cdots \oplus R_k$ of rings $R_i$ then $R^*$ decomposes naturally as
a direct product $R^*=R^*_1\oplus \cdots \oplus R^*_k$ of groups.
\end{lem}

So let $\lambda\in R^*$ a unit in $R$ then $\lambda=(\lambda_1,\lambda_2,\ldots,\lambda_k) $
where each $\lambda_i\in R_i^*$
\begin{rem} If   $\prod R_{i=1}^k$ is a  direct
decomposition of a finite principal ideal ring $R$ and $\lambda \in R^*$ then
 $\lambda= CRT(\lambda_1,\lambda_2,\ldots,\lambda_k)$ where each $\lambda_i\in R_i^*$.
\end{rem}

\begin{lem}
\label{lem:2} Let $R$ be a finite principal ideal rings and $\prod R_{i=1}^k$ its direct decomposition (ie $R= CRT(R_1,R_2,\ldots,R_k)$).\\
$R$ has units $\lambda$ and $\delta$ such that $\lambda=\delta^n$ if and only if each finite chain ring $R_i$ has units $\lambda_i$ and $\delta_i$ such that $\lambda_i=\delta_i^n$.
\end{lem}
\pf
If there exist units $\lambda_i,\delta_i \in R_i$ such that $\lambda_i=\delta_i^n$ for $1\leq i\leq k$.Then
$\lambda= CRT(\lambda_1,\lambda_2,\ldots,\lambda_k)$ and $\delta =CRT(\delta_1,\delta_2,\ldots,\delta_k)$
satisfies $\lambda=\delta^n$.\\
If $R$ has units $\lambda$ and $\delta$ such that $\lambda=\delta^n$ then $\lambda_i=\psi_i(\lambda)=\psi_i(\delta^n)=\psi_i(\delta)^n=\delta_i^n$
\qed

%\begin{lem}
%\label{lem:2} Let $R$ be a finite principal ideal rings and $\prod R_{i=1}^k$ its direct decomposition (ie $R= CRT(R_1,R_2,\ldots,R_k)$).\\
%$R$ has units $\lambda$  such that $\lambda^2=1$ if and only if each finite chain ring $R_i$ has units $\lambda_i$ such that $\lambda_i^2=1$.
%\end{lem}
%\pf
%If there exist units $\lambda_i\in R_i$ such that $\lambda_i^2=1$ for $1\leq i\leq k$.\\
%Then $\lambda= CRT(\lambda_1,\lambda_2,\ldots,\lambda_k)$ satisfies $\lambda^2=1$.\\
%If $R$ has a unit $\lambda$ such that $\lambda^2=1$ then $\lambda_i^2=\psi_i(\lambda)^2=\psi_i(\lambda^2)=\psi_i(1)=1$
%\qed

\begin{thm}
\label{thm:prod} Let $R$ be a finite principal ideal ring, $\prod
R_{i=1}^k$ its direct decomposition and $\lambda$ be a unit in $R$
such that $\lambda= CRT(\lambda_1,\lambda_2,\ldots,\lambda_k)$ with
$\lambda_i\in R_i^*$. Let $C=CRT(C_1,C_2,\dots,C_k)$ be a code over
$R$ of length $n$ with local components codes $C_i$ of length $n$
over $R_i$; $1\le i\le k$ be codes of length an integer $n$. Then
$C$ is $\lambda$-constacyclic code over $R$ if and only if each
$C_i$ is $\lambda_i$-constacyclic code over $R_i$.
\end{thm}
\pf
 For $i\in \{1,\ldots, k\}$ let $\F_{q_i}$ be the residual fields of
 $R_i$.
 %such that $(n,q_i)=1 \,\forall i \in\{1,2,\ldots,k\}$.
 Further,define the following ring homomorphism
\begin{tabular}{cccc}
               $ \phi $: $ R[x]/\langle x^n-\lambda \rangle$ &$ \longrightarrow$ & $R_i[x]/\langle x^{n}-\lambda_i\rangle$ \\
                     $a_0+a_1x+\cdots a_{n-1}x^{n-1} $& $\longmapsto $& $\psi_i(a_0)+\psi_i(a_1)x + \cdots +\psi_i(a_{n-1})x^{n-1}$ \\
                \end{tabular}

Next define
\[
\phi :R[x]/(x^n-\lambda)\longrightarrow R_1[x]/(x^n-\lambda_1)\times R_2[x]/(x^n-\lambda_2)\times\cdots \times R_k[x]/(x^n-\lambda_k)
\]
where
\[
\phi(f(x))= (\phi_1(f(x)),\phi_2(f(x)),\cdots, \phi_k(f(x))).
\]

If $I$ is an ideal of $R[x]/(x^n-\lambda)$, then $\phi_i(I)$ is an ideal of $R_i[x]/(x^n-\lambda_i)$.\\

Conversely for ideals $I_i$ in $R_i[x]/(x^n-\lambda)$ we define
\[
\phi^{-1}(I_1,I_2,\ldots,I_k).
\]
Note that
\[
I=CRT(I_1,I_2,\dots,I_k)
\]
is the unique ideal in $R[x]/(x^n-\lambda)$  that is congruent to $I_i$ in $R_i$.
By the generalized Chinese Remainder Theorem this map is well defined,
and furthermore
\[
I=CRT(I_1,I_2,\dots,I_k)
\]
is an ideal in $R[x]/(x^n-\lambda)$.
Associating a cyclic code with its corresponding ideal we have that
\[
CRT(C_1,C_2,\ldots,C_k)
\]
is $\lambda$-constacyclic code over $R$ if and only if each $C_i$ is $\lambda_i$-constacyclic code over $R_i$.
\qed
\begin{cor}With the above assumptions $R[x]/(x^n-\lambda)$ is principal ring if and only if
 $R_i[x]/(x^n-\lambda_i)$ is principal ideal for all $1\leq i\leq k$.
 \end{cor}
 \pf
Let $C$ a $\lambda$constacyclic code of length $ n$ over $R$,generated by $f(x)\in R[x]/(x^n-\lambda)$ then by Theorem \ref{thm:prod}
and since $C=CRT(C_1,C_2,\ldots,C_k)$ then $C_i$ is generated by $\phi_i(f(x))$ which a polynomial in $R_i[x]/(x^n-\lambda_i)$.
So $C_i$ is principal.
Conversely,let $C_i$ be a cyclic codes of length $n$ over $R_i$ generated by $f_i(x)\in R_i[x]/(x^n-\lambda_i)$,
Let $f(x) \in R[x]/(x^n-\lambda)$ such that $f(x)= \phi^{-1}(f_1(x),f_2(x),\cdots,f_k(x))$.
 Since $\Phi$ is a ring isomorphic the $f(x)$ is unique.
 Let $D$ the cyclic code generated by $f(x)$ then
 \[
D= CRT(C_1,C_2,\ldots,C_k)
\]
 Or the Chinese Remainder Theorem $CRT(C_1,C_2,\ldots,C_k)$ is unique thus $C=D$
 \qed
 \begin{rem}
 Let $R$ be a finite principal ideal ring, $\prod
R_{i=1}^k$ its direct decomposition,$\F_{q_i}$ the residue field of each $R_i$
such  that $(n,q_i)=1 \,\forall i \in\{1,2,\ldots,k\}$.
and let$\lambda= CRT(\lambda_1,\lambda_2,\ldots,\lambda_k)$
 By Theorem \ref{th:prince}
each $R_i[x]/(x^n-\lambda_i)$ is principal ideal ring, thus $ R[x]/(x^n-\lambda)$ is principal ideal ring.
If there exist $i\in\{1,\ldots,k\}$ such that  $R_i$ is a field $R_i[x]/(x^n-\lambda_i)$ is a principal ideal rind
for all length.
 \end{rem}
\begin{ex}
Let $\F_p$ be the finite field with $p$ a prime elements and $R=\F_p[x]/(v^2-v)=\F_p+v\F_p$.
Since $\langle v\rangle$ and $\langle (1-v)\rangle$ are the unique ideal maximal of index
of stability 1.Then  $(v),(1-v)$ is the  direct
decomposition of $R$.
Note that any element $c$ of $R^n$ can be expressed
as $c=a+vb$ = $v(a+b)+(1-v)a$ where $a,b \in \F^n_p$.
 Let:
\begin{equation}
\begin{array}{ccl}
\label{eq:doug}\psi:R^n&\longrightarrow & \F_p^n\times \F_p^n \\
 a+bv &\mapsto & \psi(a+bv) = (\psi_1(a+bv),\psi_2(a+bv))=(a+b,a)
\end{array}
\end{equation}
be the canonical $R$-module isomorphism.
and
for $i=1,2$, let $C_i$ be a code over $\F_p$ of length $n$ and
let
\[
C=CRT(C_1,C_2)=\Psi^{-1}(C_1\times C_2)= \{\Psi^{-1} (\vv_1,\vv_2) \ |\ \vv_1 \in C_1,\vv_2 \in C_2 \}.
\]
We refer to $C$ as the \textit{Chinese product of codes $C_1,C_2$} \cite{MDR}.
By Theorem\ref{thm:prod} a $\lambda$-constacyclic code over $R$ if and only if each $C_i$ is $\lambda_i$-constacyclic code over $\F_p$.
with $\lambda=CRT(\lambda_1,\lambda_2)$.
Let $\lambda=1-2v=-v+(1-v)$ so $\lambda=CRT(-1,1)$
By Theorem\ref{thm:prod} any $(1-2v)$-constacyclic code $C$ over $R$ is the Chinese Remainder Theorem
 of a negacyclic code $C_1$ over $\F_p$ and a cyclic code $C_2$ over $\F_p$ such that $C=CRT(C_1,C_2)$.
\end{ex}
These codes have also been studied by~~\cite{Zhu}.

Let $n$ be an integer and $\lambda_i,\delta_i$  units in $R_i$
such that $\lambda_i=\delta^n$. Let:\\
 $\mu_i :R_i[x]/\langle
x^n-1\rangle \mapsto R_i[x]/ \langle x^n- \lambda \rangle$ defined by
$\mu_i(c(x))=c(\delta_i^{-1} x)$.\\
  By Proposition \ref{prop:isom}  we have that $\mu_i$ is a ring
isomorphism.

In the following we generalize the result above to finite principal ideals rings.
\begin{prop}
\label{prop:isomprincipal} Let $n$ be an  integer and $\lambda = CRT(\lambda_1,\ldots,\lambda_k)$ and $\delta = CRT(\delta_1,\ldots,\delta_k)$  units
such that $\lambda=\delta^n$.
Let
\begin{equation}
\begin{array}{ccl}
\label{eq:ling2}\mu:R[x]/\langle x^n-1\rangle &\longrightarrow & R/\langle x^n-\lambda \rangle\\
 c(x)&\mapsto & \mu(c(x))=(\mu_1(c(x),\ldots,\mu_k(c(x)))=(c(\delta_1^{-1} x),\ldots,c(\delta^{-1}_k x)).
\end{array}
\end{equation}
Then  $\mu$ is a ring isomorphism.
\end{prop}
\pf
We can proof easily that $\mu$ is a ring  isomorphism,
since $R[x]/\langle x^n-1\rangle \simeq \Pi_{i=1}^k R_i[x]/\langle x^n-1\rangle$ and  by lemma \ref{lem:2} we deduce  that
$\lambda=\delta^n \Longleftrightarrow \lambda_i=\delta_i^n$, $\forall i\in\{1,\ldots,k\}$ then by Proposition \ref{prop:isom}
$\Pi_{i=1}^k R_i[x]/\langle x^n-1\rangle\simeq\Pi_{i=1}^k R_i[x]/\langle x^n-\lambda_i\rangle$ , $\forall i\in\{1,\ldots,k\}$
and  $R[x]/\langle x^n-\lambda\rangle \simeq \Pi_{i=1}^k R_i[x]/\langle x^n-\lambda_i\rangle$ then
we obtain the result
\qed
Since $(n,q_i)=1$ with $\F_{q_i}$ the residue field of the finite chain ring  $R_i$, then $R_i[x]/
\langle x^n- \lambda_i \rangle$ is principal ring $\forall i\in\{1,\ldots,k\}$ then the ideals in $R[x]/
\langle x^n- \lambda \rangle$ are principal ideals. So the
following result is a straightforward consequence of
Proposition~\ref{prop:isomprincipal}.
\begin{cor}
\label{cor:squareprincipal}
Let $R$ be a finite principal ideal  ring and $\lambda,\delta$  units in $R$ such that $\lambda=\delta^n$ . A subset
$I$ in $R[x]$ is an ideal in $R[x]/\langle x^n-1\rangle$ if and only
if $\mu(I)$ is an ideal in $R[x]/\langle x^n-\lambda \rangle$.
Equivalently, the set $C$ is a cyclic code of length $n$ over the
principal ideal ring $R$ if and only if $\mu(C)$ is a $\lambda$-constacyclic
code of length $n$ over $R$.
\end{cor}
\begin{ex}
Let  $R=\F_p[x]/(v^2-v)\simeq\F_p+v\F_p$,
  $n$ an odd integer  we deduce by \ref{prop:isomprincipal} that any $(1-2v)$-constacyclic code over
 $R$ is isomorphic to a cyclic code over $R$.Thus $C_1$ and $C_2$ are cyclic codes over $\F_p$.
 Conversely if $n$ is odd integer then any negacyclic code $C_1$ over $\F_p$ is equivalent to cyclic code over $\F_p$
 and so by Theorem \ref{thm:prod} $CRT(C_1,C_2)$ is cyclic code over $R$.
 \end{ex}
\section{Self-dual Cyclic Codes over Finite Ideal Principal Rings}
Since any finite ideal principal ring is a direct product of some
finite chain rings, one starts by giving some results on the latter.

\subsection{Cyclic Self-dual Codes over Finite Chain Rings}

In this subsection, we consider cyclic self-dual codes over finite
chain rings. For a polynomial $f(x)$ of degree $r$, let
$f^{\star}(x)$ denote its reciprocal polynomial $x^{r}f(x^{-1})$.
The following lemma is easy to obtain.
\begin{lem}
\label{lem:reci}Let $f(x)$ and $g(x)$ be two polynomials in $R[x]$
with $\deg f(x) \ge \deg g(x)$. Then the following holds.
\begin{itemize}
%\item[(i)] $[f(x)g(x)]^*=f(x)^*g(x)^*$.
\item[(i)] $[f(x)+g(x)]^*=f(x)^*+x^{\deg f- \deg g }g(x)^*$.
\item[(ii)] If $f$ is monic, then $\overline{f^*}=\overline{f}^*$.
\end{itemize}
\end{lem}

The following theorem gives the structure of the dual of a cyclic
code over a finite chain ring.
\begin{thm}
\label{th:perdual} (\cite[Theorem 3.8]{permounth}) Let $R$ be a
finite chain ring with characteristic $p$, maximal ideal $\gamma$,
and index of nilpotency $e$. Let $n$ be an integer such that
$(p,n)=1$ and $f_{1}f_{2}\ldots f_{l}$ be the representation of
$x^{n}-1$ as a product of basic irreducible pairwise coprime
polynomials in $R[x]$. If $C$ is a cyclic code of length $n$  over
$R$, then $C^{\bot} =
\langle\hat{F}_{0}^{*},\gamma\hat{F}_{e}^{*}, \ldots,
\gamma^{e-1}\hat{F}_{2}^{*}\rangle$, where $F_0,F_1,\ldots,
F_{e-1}$ are pairwise coprime polynomials which are divisors of
$x^n-1$ as given in Theorem~\ref{th:prince}.
\end{thm}

\begin{prop}(\cite[Proposition 4.3]{permounth}) \label{thm:main}
Let $R$ be a finite chain ring with even index of nilpotency $e$ and
maximal ideal $\gamma$. Then there exists a non-trivial self-dual
cyclic code over $R$ if and only if there exists a basic irreducible
factor $ f\in R[x]$ of $x^{n} - 1$ such that $f$ and $f^{*}$ are
not associate.
\end{prop}
The following Theorem was given first by Kanwar and L\'{o}pez-Permouth \cite{K-L} and after
 by  H. Dinh and S. R. L\'{o}pez-Permouth \cite{permounth}
but with a false proof,so we give it again  with another proof.
In \cite{permounth ,K-L} the authors proof that all  cyclotomic cosset  mod $n$ must be
reversible for having  the congruence  $(p^r)^i\equiv -1 \pmod n$ for a positive integers $i$.
While we need only to have the first cyclotomic cosset $C_1$ to be reversible.
But before that  we give this useful lemma.
\begin{lem}
\label{lemma reversible}
\[  C_1 \text{ is reversible} \Longrightarrow \forall j \in \mathbb{Z}_n, C_j \text{ is reversible}\]
\end{lem}
\pf
If $C_ 1$ is reversible, then there exit  $k$, $1 \leq k
\leq \ord _n (q)$ such that
$q^k \equiv  -1  \mod n,$ which means that  $j{q^k} \equiv -j \mod n$,
then $C_j=C_{-j}$.
\qed
\begin{thm}
\label{th:nonTriv} Let $R$ be a finite chain ring with maximal ideal
$\gamma$, index of nilpotency $e$ even, and residue field $K$ where
$|R|=p^{er}$ and $|K|=p^r$. Then non-trivial cyclic self-dual codes
of length $n$ over $R$ exist if and only if $(p^r)^i\neq -1 \pmod n$
for all positive integers $i$.
\end{thm}
\pf Let $f(x)$ be a monic basic irreducible polynomial which divides $x^n-1$.
Then $\overline{f(x)}$ is a minimal irreducible polynomial over $F_q[x]$.
Hence there exists a cyclotomic class $C_u$ associated with $\overline{f(x)}$.
Therefore $\overline{f(x)}=\prod_{i\in C_u}(x-\alpha^i)$, where $\alpha$ is a primitive $n$th root of unity.
The reciprocal polynomial of $\overline{f(x)}$ is the polynomial
$\overline{f(x)}^*=(\prod_{i\in C_u}(x-\alpha^i))^*=x^r \prod_{i\in C_u}(x^{-1}-\alpha^i)=\prod_{i\in C_{n-u}}(x-\alpha^i)$,
but by Lemma~\ref{lem:reci} we have $\overline{f(x)^*}=\overline{f(x)}^*$.

By Theorem~\ref{thm:main}, a non-trivial cyclic self-dual code exists if
and only if there is a basic irreducible polynomial $f(x)$ a factor of
$x^n-1$ such that $f(x)$ and $f(x)^*$ are not associated.
We show that this can occur if and only if $(p^r)^i\neq -1 \pmod n$ for all positive integers $i$.

Let $\bar{f}(x)\in F_q[x]$ be irreducible and $f(x)/x^n-1$.
Then $\bar{f}(x) =\prod_{i\in C_u}(x-\alpha^i)$ where $C_u$ is
the cyclotomic coset of $n$ that contains the smallest element $u$ and $\alpha $ is a primitive $n$-th root of unity .
Now if $(p^r)^i\neq -1 \pmod n$ for all positive integers $i$, then $C_1\neq C_{-1}$.
Hence $(f(x))\neq (f^*(x))$ where
$\bar{f}(x)= \prod_{i\in C_1}(x-\alpha^i)$, and the code $(fg,\gamma ^{\frac{e}{2}}ff^*)$ is a non-trivial self-dual code where $ff^*g= x^n-1$.
Conversely, if a non-trivial cyclic self-dual code exists then by \ref{thm:main} there exists a factor $f(x)/x^n-1$ with
$(f(x))\neq (f(x)^*)$.
Hence $C_u\neq C_{-u}$, and then by Lemma \ref{lemma reversible} $C_1\neq C_{-1}$ where $\bar{f(x)} =\prod_{i\in C_u}(x-\alpha^i)$.
Therefore $(p^r)^i\neq -1 \pmod n$ for all positive integers $i$, because otherwise $C_u = C_{-u}$ for all cyclotomic
cosets and $(f(x))=(f(x)^*)$ for any $f(x)/x^n-1$.
\qed

\begin{lem}
\label{lem:skersys} Let $n$ and $s$ be positive integers, and $q$ a
prime power. Then the following holds.
\begin{itemize}
\item[(i)] If $q^s\equiv -1 \bmod n$, then $\ord_n(q)$ is even.
\item[(ii)] If $n$ is prime, then we have $\ord_n(q)$ is even if and only if $\exists i$ such that $q^i \equiv-1 \bmod n$.
\end{itemize}
\end{lem}
\pf Part (i) follows from \cite[Proposition 4.7.5]{skersys}. For Part (ii),
assume that $\ord_n (q)=2w$ is even, so then $q^{2w}\equiv 1\bmod
n$. Hence $n|(q^{w}-1)(q^{w}+1)$. Since $n$ is prime and cannot
divide $q^{w}-1$ (because of the order), we have that $q^{w}=-1
\bmod n$. The converse follows from Part (i).
\qed

%For the binary case, all integer $n$ such that $2^i\neq -1 (\mod n)$ are known~\cite{moree}.
%Conversely, less is known for $p$ odd.
The following result answers the question posed in~\cite[p.
1734]{permounth} by providing a simple criteria for the existence of
cyclic self-dual codes.
\begin{thm}
\label{thm:ord} Let $R$ be a finite chain ring with maximal ideal
$\gamma$, index of nilpotency $e$ even, and $|R|=p^{er}$, where
$|K|=p^r$. Then non-trivial cyclic self-dual codes of odd length $n$
a power of a prime exist over $R$ if and only if $\ord_n(p^r)$ is odd.
\end{thm}
\pf If there are no non-trivial self-dual codes, then by
Theorem~\ref{th:nonTriv} there exists an integer $i$ such that
$(p^r)^i \equiv -1 \bmod n$. Then by Part (i) of
Lemma~\ref{lem:skersys}, we have that $\ord_n(p^r)$ is even.

Conversely, assume that there exists a non-trivial cyclic self-dual
code. Then from Theorem~\ref{th:nonTriv} there is no integer $i$
such that $p^{ri} \equiv -1 \bmod n$. We need to show that in this
case $\ord_n(p^r)$ is odd. For this, consider the following cases.
\begin{itemize}
\item[(i)] If $n$ is an odd prime, then by Part (ii) of
Lemma~\ref{lem:skersys}, we have $\ord_n(p^r)$ is odd.

\item[(ii)] For $n =q^{\alpha}$, assume that $\ord_{q^{\alpha}}(p^r)$ is even.
We first must prove the implication
\[
\ord_{q^{\alpha}}(p^r) \text{ is even}\Rightarrow\ord_q(p^r) \text{ is even.}
\]
Assume $\ord_{q^{\alpha}}(p^r)$ is even and $\ord_q(p^r)$ is odd.
Then there exist $i>0$ odd such that $p^{ri}\equiv 1 \mod  q
 \Leftrightarrow p^{ri}=1+kq.$ Hence $p^{riq^{\alpha -1}}=(1+kq)^{q^{\alpha- 1}} \equiv 1\mod q^{\alpha}$,
because $(1+ kq)^{q^{\alpha -1}} \equiv 1+kq^{\alpha}\mod
q^{(\alpha+1}$ (the proof of the last equality can  be found in
\cite[Lemma~3.30]{demazure}). Hence
\begin{eqnarray}
\label{fermat} p^ {riq^{\alpha -1}} \equiv 1\mod q^{\alpha}.
\end{eqnarray}
If we have $i$ odd and ${q^{\alpha -1}}$ odd, then
$\ord_{q^{\alpha}}(p^r)$ is odd (because $\ord _{q^{\alpha}}(p^r) |
i q^{{\alpha}-1}$), which is absurd. Hence $\ord_q(p^r)$ is even, so
there exists some integer $j$ such that $0<j< \ord_q (p^r)$, and
$p^{rj}\equiv -1\mod q$. Then from (\ref{fermat}) we have
$p^{rjq^{\alpha -1}}\equiv -1\mod q^{\alpha}$. This gives that the
cyclotomic class $C_1$ is reversible, which by
Theorem~\ref{th:nonTriv} is impossible.
\end{itemize}
\qed

\begin{rem}
Note that $\ord_n(p^r)$ odd is a sufficient condition for all $n$ for the existence of
a self-dual code over $R$.
\end{rem}

For the remainder of the paper, the notation $q=\square \bmod n$
means that $q$ is a residue quadratic modulo $n$.
\begin{cor}
\label{cor:linea} Let $R$ be a finite chain ring with maximal ideal
$\gamma$, index of nilpotency $e$ even, and residue field $K$ such
that $|K|=p^r$. Then if $p_1\ldots p_s$ is the prime factorization
of an odd integer $n$ such that $p^r=\square \bmod p_i$ and $p_i
\equiv -1 \bmod 4$ for $1\leq i \leq s$, then there exists a
non-trivial cyclic self-dual code over $R$.
\end{cor}
\pf We have that $\ord_n(p^r)=\mbox{lcm}(\ord_{p_i}(p^r))$. Since
$p^r=\square \bmod p_i$, then $\ord_{p_i}(p^r)$ divides
$\frac{p_i-1}{2}$. Hence $\ord_{p_i}(p^r)$ is odd, otherwise $p_i
\equiv 1 \bmod 4$. Then  $\ord_n(p^r)$ is odd, and by
Theorem~\ref{thm:ord} we have the existence of a non-trivial cyclic
self-dual code. \qed
\begin{cor}
\label{cor:prime}
With the previous notation, if  $n$ is an odd prime such that
$n\equiv -1 \mod 4$, then there exists a cyclic self-dual code if
and only if $p=\square \mod n$.
\end{cor}
\pf The necessary condition is given by~\cite[Corollary
4.7]{permounth}. For the converse, if we assume $p=\square \mod n$,
then $p^r=\square \mod n$, and the result follows from
Corollary~\ref{cor:linea}. \qed

%\begin{defi}
%A code $C$ over a finite chain ring $R$ with residue field $K$ is
%said to be the lift of a code $C'$ over $K$ if $\overline{C}=C'$,
%where $-$ denotes reduction modulo $\gamma$ as given in
%(\ref{eq:over}).
%\end{defi}
For a cyclic code of length $n$ with $(n,p)=1$, using
Theorems~\ref{lem:aicha} and~\ref{th:prince} and Hensel's Lemma, we
have the following result.
\begin{thm}~(\cite[Theorem 4.20]{G-G})
\label{th:free} Let $C$ be a cyclic code of length $n$ over a finite
chain ring $R$ with characteristic $p$ such that $(p,n)=1$. Then $C$
is a free cyclic code with rank $k$ if and only if there is a
polynomial $f(x)\in R[x]$ such that $f(x)|(x^n-1)$ generates $C$. In
this case, we have $k=n-deg(f)$.
\end{thm}
\begin{thm} \cite{aicha2011}
\label{th:odd} Let $R$ be a finite chain ring with maximal ideal
$\langle\gamma\rangle$, index of nilpotency $e$ , and characteristic
$p$. Then if $p$ is odd, there is no free cyclic self-dual code of
length $n$ over $R$ with $(p,n)=1$.
\end{thm}

\subsection{Self-dual Cyclic Codes over Principal Rings}

If $R$ is a finite principal ideal ring, we say that the
decomposition of $R$ into a product of finite chain rings, as in
(ii), is a \textit{canonical decomposition of $R$}. The ideal
$\mathfrak m_1, \mathfrak m_2,\dots , \mathfrak m_k$ in this case is
called a \textit{direct decomposition of $R$}.

%Let $\mathfrak m_1, \mathfrak m_2,\dots , \mathfrak a_k$ a such
%direct decomposition
\bigskip
Let $R$ be a finite ring and $(\mathfrak{a}_i)_{i=1}^n$  a direct
decomposition of $R$. Let $\Psi: R^n \rightarrow \prod_{i=1}^k
R_i^n$ be the canonical $R$-module isomorphism.
For $i=1,\dots, k$, let $C_i$ be a code over $R_i$ of length $n$ and
let
\[
C=CRT(C_1,C_2,\dots,C_k)=\Psi^{-1}(C_1\times \dots \times C_k)= \{\Psi^{-1} (\vv_1,\vv_2,\dots,\vv_k) \ |\ \vv_i \in
C_i \}.
\]
We refer to $C$ as the \textit{Chinese product of codes $C_1,C_2,\dots,C_k$} \cite{MDR}.

\begin{thm}
\label{thm:prod2} With the above notation, let $C_1,C_2, \ldots ,C_k$
be codes of length $n$ with $C_i$ a code over $R_i$, and let
$C=CRT(C_1,C_2,\dots,C_k)$. Then we have the following.
\begin{itemize}
\item[(i)]  $C$ is a cyclic code if and only if each $C_i$ is a cyclic code
\item[(ii)] $C_1,C_2,\dots,C_k$ are self-dual codes if and only if $C$ is a self-dual code.
\end{itemize}
\end{thm}
\pf
The part (i) is a particular case of Theorem \ref{thm:prod}.

\[
CRT(C_1,C_2,\ldots,C_k)^{\perp}=CRT(C_1^\perp,C_2^\perp,\ldots,C_k^\perp).
\]
Then if $C= CRT(C_1,C_2,\ldots,C_k)$ we have that
\[
C^\perp =CRT(C_1^\perp,C_2^\perp,\ldots,C_k^\perp)= CRT(C_1,C_2,\ldots,C_k)=C,
\]
and the code $C$ is self-dual.
\qed

In the following we generalize the theorem  \ref{thm:ord} to  finite principal ideal rings.
\begin{thm}
 Let  $R\cong \prod_{i=1}^k R/\mathfrak m_i^{t_i}=\prod_{i=1}^k R_i$, be a finite
principal ideal ring,$\F_{q_i}$ the residue field of $R_i$ for $1\leq i\leq k$ and $C$ a cyclic code over $R$.
Then $C$ is self-dual code of length a power of a prime odd $n$ if and only if $\ord_n(q_i)$ is odd for $1\leq i\leq k$.
\end{thm}
\pf Let $n$ a power of a prime odd such that $(n,q_i)=1$ and $C=CRT(C_1,C_2,\dots,C_k)$ a cyclic self-dual code over $R$ then by Theorem \ref{th:odd}
$C_i$ is a cyclic self-dual code over $R_i$ for all $1\leq i\leq k$ and by Theorem \ref{th:odd} $\ord_n(q_i)$ is odd.\\
Conversely if $\ord_n(q_i)$ is odd then there exist a cyclic self-dual code $C_i$ over $R_i$ for all $1\leq i\leq k$,
then by Theorem \ref{thm:prod} the cyclic code $C=CRT(C_1,C_2,\dots,C_k)$ is self-dual cyclic code over $R$.
\qed

For the remainder of the paper, the notation $q=\square \bmod n$
means that $q$ is a residue quadratic modulo $n$.
%\begin{cor}
%\label{cor:linea} Let $R$ be a finite chain ring with maximal ideal
%$\gamma$, index of nilpotency $e$ even, and residue field $K$ such
%that $|K|=p^r$. Then if $p_1\ldots p_s$ is the prime factorization
%of an odd integer $n$ such that $p^r=\square \bmod p_i$ and $p_i
%\equiv -1 \bmod 4$ for $1\leq i \leq s$, then there exists a
%non-trivial cyclic self-dual code over $R$.
%\end{cor}
In the following we generalize the corollary   \ref{cor:linea} to  finite principal ideal rings.
\begin{cor}
Let  $R\cong \prod_{i=1}^k R/\mathfrak m_i^{t_i}=\prod_{i=1}^k R_i$, be a finite
principal ideal ring,$\F_{q_i}$ the residue field of $R_i$ , $n$ an integer such that $(n,q-i)=1$
for $1\leq i\leq k$.Then if $p_1\ldots p_s$ is the prime factorization
of an odd integer $n$ such that $q_i=\square \bmod p_j$ and $p_j
\equiv -1 \bmod 4$ for $1\leq j \leq s$, then there exists a
non-trivial cyclic self-dual code over $R$
\end{cor}
\pf
If  if $n=p_1\ldots p_s$ such that $q_i=\square \bmod p_j$ and $p_j \equiv -1 \bmod 4$ for $1\leq j \leq s$
By corollary \ref{cor:linea} there exist exists a
non-trivial cyclic self-dual code $C_i$ over $R_i$.
 Then by Theorem \ref{thm:prod} the cyclic code $C=CRT(C_1,C_2,\dots,C_k)$ is self-dual cyclic code over $R$.
\qed

%\begin{cor}
%\label{cor:prime}
%With the previous notation, if  $n$ is an odd prime such that
%$n\equiv -1 \mod 4$, then there exists a cyclic self-dual code if
%and only if $p=\square \mod n$.
%\end{cor}
In the following we generalize the corollary  \ref{cor:prime} to  finite principal ideal rings.
\begin{cor}
With the previous notation, if  $n$ is an odd prime such that
$n\equiv -1 \mod 4$, then there exists a cyclic self-dual code if
and only if $p_j=\square \mod n$,where $q_j=p_j^r$.
\end{cor}
\pf
Let  $n$ is an odd prime such that $n\equiv -1 \mod 4$
If $p_j=\square \mod n$ then by corollary \ref{cor:prime} there exist a self-dual cyclic code $C_j$
of length $n$ over $R_j$. Then by Theorem \ref{thm:prod}
the cyclic code $C=CRT(C_1,C_2,\dots,C_k)$ is self-dual cyclic code over $R$.
\qed
In the following we generalize the theorem  \ref{th:odd} to  finite principal ideal rings.
%\begin{thm}
%\label{th:odd} Let $R$ be a finite chain ring with maximal ideal
%$\langle\gamma\rangle$, index of nilpotency $e$, and characteristic
%$p$. Then if $p$ is odd, there is no free cyclic self-dual code of
%length $n$ over $R$ with $(p,n)=1$.
%\end{thm}
\begin{thm}
Let  $R\cong \prod_{i=1}^k R/\mathfrak m_i^{t_i}=\prod_{i=1}^k R_i$, be a finite
principal ideal ring,$\F_{q_i}$ the residue field of $R_i$ , $n$ an integer such that $(n,q-i)=1$
for $1\leq i\leq k$ and $C=CRT(C_1,C_2,\dots,C_k)$ a cyclic code over $R$,
If there exist $i\in \{1,\ldots,k\}$ such that $q_i$ is odd and $C_i$ is free then $C$ is not self-dual.
\end{thm}
\pf Let $C=CRT(C_1,C_2,\dots,C_k)$ a cyclic code of length $n$ over $R$ such that $(n,q_i)=1$  for $1\leq i\leq k$ then
By theorem \ref{th:odd} if $q_i$ is odd and $C_i$ is free ten $C_i$ can not be self-dual,so by Theorem \ref{thm:prod}
$C$ can not be self-dual cyclic code of length $n$ over $R$.
\qed

\section{Cyclic Codes over Finite Ideal Principal Rings with Odd Index of Stability}

In this section, we prove that there is no simple root cyclic self-dual codes over finite chain rings when the nilpotency index
of the generator of the maximal ideal is odd and we generalize it to finite ideal principal rings when the stability index
of the generator of one of the maximal ideals is odd.

\begin{thm}
\label{th:none} Let $R$ be a finite chain ring where $\langle
\gamma \rangle $ is the maximal ideal with nilpotency index $e$. If
$e$ is odd and $q$ a prime power then there are no nontrivial
self-dual cyclic code of length $n$ over $R$ such that $(n,q)=1$.
\end{thm}
\pf If $q=2^k$, then $(n,q)=1$ and $n$ must be odd, so that from~Remark~\ref{rem:wood} $e$ must be even.
%If $C$ is a cyclic code over $R$, then its dual is also a cyclic code over $R$.
Let $C$ be a non-trivial cyclic code of length $n$ over $R$ so there exists
monic and coprime polynomials $F_0,F_1,\ldots, F_{e-1},F_{e}$ such
that $x^{n}-1=F_0F_1\ldots F_{e-1}F_{e}$ and $C=\langle\hat{F}_1,\gamma \hat{F}_2, \ldots,\gamma^{e-1} \hat{F}_e \rangle$.
If $C$ is self-dual, then from~\cite[Proposition 4.1]{permounth} $F_{i}$ is associate with $F_{j}$ for $i,j \in \{ 0,1,\ldots e\}$ and
$ i+j \equiv 1 \pmod{e+1}$.
Then $F_{i}$ = $\epsilon F_{j}^{*}$ for all $i,j \in \{0,\ldots e\}$
$ i+j \equiv 1 \pmod{e+1}$, $\epsilon$ a unit in $R$.
Then $F_{i}\neq F_{j}^{*}$ since $e$ is odd and it cannot be that $i+i\equiv e+2$, so
therefore
\[
x^{n}-1 = F_{0}F_{0}^{*}F_{2}F_{2}^{*}F_{3}F_{3}^{*}
\ldots F_{\frac{e+1}{2}}F_{\frac{e+1}{2}}^{*}.
\]
Thus none of the $F_i$ are self-reciprocal. The polynomial $(x-1)$ is a factor
of $x^{n}-1$, so there is an $0\le i_0 \le e$ such that
$F_{i_0}=(x-1)g(x)$ for some polynomial $g(x)$.
Hence
\[
F_{i_0}^{*}=(x-1)^{*}g(x)^{*}=(x-1)g(x)^{*}=F_{1-i_0 \pmod{1+e}},
\]
which is impossible since for all $0\le i \le e$ the $F_i$ are coprime,
and $x^n-1$ has no repeated roots since $(n,q)=1$.
\qed

\begin{thm}
Let $R\cong \prod_{i=1}^k R/\mathfrak m_i^{t_i}$, be a finite
principal ideal ring, and $C$ a cyclic code over $R$.
Then if one of the $t_i$ is odd, $C$ cannot be a self-dual code.
\end{thm}
\pf From Theorem~\ref{thm:prod}, $C$ is cyclic and self-dual if and
only if all $C_i$ are also cyclic and self-dual. However,
from~Theorem~\ref{th:none} if there exists an $i$ such that $t_{i}$
is odd, then the code $C_{i}$ cannot be self-dual.
\qed

\section{Constacyclic Codes over $R+vR$}

Let $R$ be a finite commutative chain ring where $\langle
\gamma \rangle $ is the maximal ideal with nilpotency index $e$
and residue field $\F_q$.
Let $R+vR=\{a+vb;\,\,a,b\in R\}$ with $v^2=v$.
this ring is a kind of finite commutative principal ideal ring.
With two coprime ideals.
$\langle v\rangle=\{av;\;\;a\in R\}$ and $\langle 1-v\rangle=\{a(1-v);\;\;a\in R\}$.
with index of stability 1 then,both $R_1=R/\langle v\rangle$ and $R_2=R/\langle 1-v\rangle$
is isomorphic to $R$.
By the Chinese Remainder Theorem,we have $R+vR \simeq R_1\times R_2 \simeq \langle v\rangle\oplus\langle1-v\rangle$.
The motivation for what we have choused this ring is that the element  $v$ and $1-v$
are nilpotent element such that $v+1-v=1$ so By Proposition\ref{prop:doug} any submodule $N$
of a module $M$ over $R+vR$ is a direct decomposition of $N_1\oplus N_2$ where $N_1=vN$ and $N_2=(1-v)N$.
In particular for a positive integer $n$, $(R+vR)^n=v(R+vR)^n\oplus(1-v)(R+vR)^n$.
Since $R+vR \simeq \langle v\rangle\oplus\langle1-v\rangle$,let $x_i \in R+vR$
such that $x_i=a_iv+b_i(1-v)$, $a_i,b_i\in R $ then $x=(x_1,x_2,\ldots,x_n)=(a_1v+b_1(1-v),a_2v+b_2(1-v),\ldots,a_nv+b_n(1-v))\in(R+vR)^n$
then $x=v(a_1,a_2,\ldots,a_n)+(1-v)(b_1,b_2,\ldots,b_n)\in vR^n\oplus(1-v)R^n$
so $(R+vR)^n=vR^n\oplus(1-v) R^n$.

Let $C$ a code of length $n$ over $R+vR$ since $C$ is a submodule of $(R+vR)^n$ over $R+vR$ such that
\[
C=CRT(C_1,C_2)=\Psi^{-1}(C_1, C_2)= \{\Psi^{-1} (\vv_1,\vv_2) \ |\ \vv_1 \in C_1 \ \vv_2 \in C_2 \}.
\]
where $C_1$ and $C_2$ are codes of length $n$ over $R$
 since the  idenpotent $v$ and $1-v$
satisfies $1+1-v=1$ then by Proposition\ref{prop:doug} $C=vC+(1-v)C\simeq C_1\times C_2$
 We use the same proof as for $(R+vR)^n=vR^n\oplus(1-v) R^n$ for having
 $vC\simeq vC_1$ and $(1-v)C\simeq(1-v)C_2$.

\begin{thm}
Let $\lambda=CRT(\lambda_1,\lambda_2)=\lambda_1v+\lambda_2(1-v)$ a unit in $R+vR$ such that $\lambda_1,\lambda_2$ are units in $R$.
Let $C$ be a linear code of length $n$ an integer over $R+vR$,
Then $C$ is a $\lambda$-constacyclic code over $R+vR$ if and only if $C_1$ and $C_2$ are  a $\lambda_1$-constacyclic code and a $\lambda_2$-constacyclic code
respectively over $R$ of length $n$
\end{thm}
\pf
It is a particular case of Theorem \ref{thm:prod}.
\qed
\begin{ex}
Let $\lambda=1-2v=-v+(1-v)$ so $\lambda=CRT(-1,1)$
By Theorem\ref{thm:prod} any $(1-2v)$-constacyclic code $C$ over $R+vR$ is the Chinese Remainder Theorem
 of a negacyclic code $C_1$ over $R$ and a cyclic code $C_2$ over $R$ such that $C=CRT(C_1,C_2)$.
\end{ex}
These codes have also been studied by~~\cite{Liao}.

In \cite{dkk} Dougherty and al gave the structure of the generator of a cyclic code of length $n$ over $\mathbb{Z}_m$
in a particular case.
In the following we give the structure of the generator of a constacyclic code over $R+vR$
in  case when the later is principal ideal.

\begin{thm}
\label{thm:generator} Let $R$ be a finite commutative chain ring where $\langle
\gamma \rangle $ is the maximal ideal with nilpotency index $e$
and residue field $\F_q$,$n$ a positive integer such that $(n,q)=1$ (If $R$ is a field we don't need this condition)
 $\lambda=\lambda_1v+\lambda_2(1-v)$ a unit in $R+vR$ such that $\lambda_1,\lambda_2$ are units in $R$.
Let $C=CRT(C_1,C_2)$ be a $\lambda$-constacyclic  code of length $n$  over $R+vR$,then there are polynomials $f_1(x),f_2(x)\in R[x]$
such that $C=\langle vf_1(x),(1-v)f_2(x)\rangle$, where $C_1=\langle f_1(x)\rangle \subseteq R[x]/(x^n-\lambda_1)$
and $C_2=\langle f_2(x)\rangle\subseteq R[x]/(x^n-\lambda_2)$.
\end{thm}
\pf
By Theorem \ref{th:prince} and since $(n,q)=1$
then the rings $R[x]/(x^n-\lambda_1)$ ,$ R[x]/(x^n-\lambda_1)$ are both principal
ideal rings,so there exist polynomials $f_1(x),f_2(x)\in R[x]$ such that $C_1=\langle f_1(x)\rangle\subseteq R[x]/(x^n-\lambda_1)$ and $ C_2=\langle f_2(x)\rangle\subseteq R[x]/(x^n-\lambda_2)$.
For any $c(x)\in C$ there exist polynomials $c_1(x),c_2(x)\in R[x]$ such that $c(x)=vc_1(x)+(1-v)c_2(x)$
then $c_1(x)\in C_1$, $c_2(x)\in C_2$,there are polynomials $k_1(x),k_2(x)\in R[x]$ such that
\[
\begin{array}{ccl}
c_1(x)&=&k_1(x)f_1(x) (mod (x^n-\lambda_1))\\
c_2(x)&= & k_2(x)f_2(x) (mod (x^n-\lambda_2))
\end{array}
\]
that means,there are  $r_1(x),r_2(x)\in R[x]$ such that
$c_1(x)=k_1(x)f_1(x)+r_1(x) (x^n-\lambda_1)$ and $c_2(x)=k_2(x)f_2(x)+r_2(x)(x^n-\lambda_2)$
Since $v(x^n-\lambda)=v(x^n-\lambda_1)$
and $(1-v)(x^n-\lambda)=(1-v)(x^n-\lambda_2)$
then
\[
\begin{array}{ccl}
c(x)&=&vc_1(x)+(1-v)c_2(x)\\
&= & v(k_1(x)f_1(x)+r_1(x) (x^n-\lambda_1))+(1-v)(k_2(x)f_2(x)+r_2(x) (x^n-\lambda_2))\\
&= &vk_1(x)f_1(x)+(1-v)k_2(x)f_2(x)+(vr_1(x)+(1-v)r_2(x))(x^n-\lambda)
\end{array}
\]
hence $vk_1(x)f_1(x)+(1-v)k_2(x)f_2(x) \mod(x^n-\lambda)$
So $c(x)\in \langle vf_1(x),(1-v)f_2(x)\rangle \subset (R+vR)/(x^n-\lambda)$
On the other hand,for any $d(x)\in\langle vf_1(x),(1-v)f_2(x) \rangle \subset (R+vR)/(x^n-\lambda)$,
there are polynomials $k_1(x),k_2(x)\in(R+vR)[x]$ such that
\[ d(x)=k_1(x)f_1(x)v+k_2(x)f_2(x)(1-v) \mod (x^n-\lambda)\]
then there are $r_1(x),r_2(x)\in R[x]$ such that $vk_1(x)=vr_1(x)$ and $(1-v)k_2(x)=(1-v)r_2(x)$
and $r(x)=vr_1(x)+(1-v)r_2(x)$
such that
 \[
\begin{array}{ccl}
d(x)&=&vd_1(x)+(1-v)d_2(x)\\
&= & vf_1(x)r_1(x)+(1-v)f_2(x)r_2(x)+r(x)(x^n-\lambda)\\
\end{array}
\]
then
 \[
\begin{array}{ccl}
vd_1(x)&=&v(f_1(x)r_1(x)+r_1(x)(x^n-\lambda_1))\\
(1-v)d_2(x)&= & (1-v)(f_2(x)r_2(x)+r_2(x)(x^n-\lambda_2))\\
\end{array}
\]
this means $d_1(x)\in \langle f_1(x)\rangle \subset R[x]/(x^n-\lambda_1)$
and $d_2(x)\in \langle f_2(x)\rangle \subset R[x]/(x^n-\lambda_2)$
hence $d_1(x)\in C_1$ ,$d_2(x)\in C_2$
then $d(x)\in C$ so $\langle vf_1(x),(1-v)f_2(x)\rangle \subset C$
this gives that $C=\langle vf_1(x),(1-v)f_2(x)\rangle$
\qed

\begin{thm}
\label{thm:generatorsum} With the above assumptions
Let $C$ be a $\lambda$-constacyclic over $R+vR$,then there is a polynomial $f(x)\in (R+vR)[x]$
such that $C=\langle f(x)\rangle$.
\end{thm}
\pf By Theorem \ref{thm:generator} there are polynomials $f_1(x)$ and $f_2(x)$ over $R+vR$ such that
$C=\langle vf_1(x),(1-v)f_2(x)\rangle$.
Let $f(x)= vf_1(x)+(1-v)f_2(x)$ obviously $\langle f(x)\rangle \subseteq C$
Note that
\[
\begin{array}{ccl}
vf(x)&=&vf_1(x)\\
(1-v)f(x)&= & (1-v)f_2(x)\\
\end{array}
\]
then hence $C=\langle f(x)\rangle$.

\subsection{Cyclic Codes over $R+vR$}
As a particular case of constacyclic codes over $R+vR$ we investigate in this subsection
cyclic codes and their duals over $R+vR$.
 Let $C=CRT(C_1,C_2)$,By Theorem \ref{thm:prod} $C$ is a cyclic code of length $n$ over $R+vR$
 if and only if $C_1$ and $C_2$ are cyclic codes of length $n$ over $R$
 and furthermore  By\ref{thm:generator} there are polynomials $f_1(x),f_2(x)\in R[x]$
such that $C=\langle vf_1(x),(1-v)f_2(x)\rangle$, where $C_1=\langle f_1(x)\rangle \subseteq R[x]/(x^n-1)$
and $C_2=\langle f_2(x)\rangle\subseteq R[x]/(x^n-1)$.
And By Theorem \ref{thm:generatorsum}  $C=\langle f(x)\rangle $ where  $f(x)= vf_1(x)+(1-v)f_2(x)$.

\begin{thm}
Let $R$ be a finite commutative chain ring where $\langle
\gamma \rangle $ is the maximal ideal with nilpotency index $e$
and residue field $\F_q$ and let $n$ an integer such that $(n,q)=1$
(If $R$ is a field we don't need this condition).
If $C=CRT(C_1,C_2)$ then
 $C^\perp=\langle vh_1(x),(1-v)h_2(x)\rangle$ where
$C_1^\perp=\langle h_1(x)\rangle$ and $C_2^\perp=\langle h_2(x)\rangle$.
\end{thm}
\pf
We know that if $C=CRT(C_1,C_2)$ then
\[
CRT(C_1,C_2)^{\perp}=CRT(C_1^\perp,C_2^\perp).
\]
So since $(n,q)=1$, the finite ring $R[x]/(x^n-1)$ is a principal ideal ring and the dual of any cyclic code is cyclic code
then there exist polynomials $h_1(x)$ and $h_2(x)$ in $R[x]$ such that
$C_1^\perp=\langle h_1(x)\rangle$ and $C_2^\perp=\langle h_2(x)\rangle$
By Theorem \ref{thm:generator} $C^\perp=\langle vh_1(x),(1-v)h_2(x)\rangle$
and By theorem \ref{thm:generatorsum} $C^\perp=\langle vh_1(x)+(1-v)h_2(x)\rangle$.
\qed
\section{Conclusions}

In this paper, the isomorphism between constacyclic codes and cyclic
codes over finite principal ideal rings was established. Necessary and
sufficient conditions were given for the existence of cyclic
self-dual codes over finite principal ideals rings.

\end{document}